\documentclass[epj]{svjour}

\usepackage{amssymb}
\usepackage{amsmath}
\usepackage{graphicx}

\begin{document}

\title{Revisiting Cardassian Model and Cosmic Constraint}

\author{Lixin Xu\inst{1,2}\thanks{lxxu@dlut.edu.cn}}

\institute{Institute of Theoretical Physics, School of Physics \&
Optoelectronic Technology, Dalian University of Technology, Dalian,
116024, P. R. China \and College of Advanced Science \& Technology,
Dalian University of Technology, Dalian, 116024, P. R. China}

\abstract{In this paper, we revisit the Cardassian model in which the radiation energy component is included. It is important for early epoch when the radiation cannot be neglected because the equation of state (EoS) of the effective dark energy becomes time variable. Therefore, it is not equivalent to the quintessence model with a constant EoS anymore. This situation was almost overlooked in the literature. By using the recent released Union2 $557$ of type Ia supernovae (SN Ia), the baryon acoustic oscillation (BAO) from
Sloan Digital Sky Survey and the WiggleZ data points, the full information of cosmic
microwave background (CMB) measurement given by the seven-year
Wilkinson Microwave Anisotropy Probe observation, we constrain the Cardassian model via the Markov Chain Monte Carlo (MCMC) method. A tight constraint is obtained: $n= -0.0479_{-    0.0732-    0.148}^{+    0.0730+    0.142}$ in $1,2\sigma$ regions. The deviation of Cardassian model from quintessence model is shown in CMB anisotropic power spectra at high $l$'s parts due to the evolution of EoS. But it is about the order of $0.1/\%$ which cannot be discriminated by current data sets. The Cardassian model is consistent with current cosmic observational data sets.}



\maketitle

\section{Introduction}

Since the discovery of current accelerated expansion of our Universe
\cite{ref:Riess98,ref:Perlmuter99}, a flood of models have been
designed to explain this late time accelerated expansion phase. For the reviews, please see
\cite{ref:DEReview1,ref:DEReview2,ref:DEReview3,ref:DEReview4,ref:DEReview5,ref:DEReview6,ref:DEReview7}. In the phenomenological perspective, the
accelerated expansion of our Universe can be realized through modifying
the form of Friedmann equation via the introduction of an extra exotic
energy component, dubbed dark energy, which has negative pressure, or
by some possible modifications of gravity theory, say $f(R)$ and brane
models etc. As a result, the conventional Friedmann equation can be
modified into the form of
\begin{equation}
H^2=f(\rho),
\end{equation}
where $f(\rho)$ is a function of energy density $\rho$ which may include dark matter and extra energy components, and $H$
is the Hubble parameter. The Cardassian model firstly proposed by
Freese and Lewis \cite{ref:Cardassian}, where the Friedmann equation
was modified into the form of
\begin{equation}
H^2=\frac{8\pi G}{3}\rho+A\rho^n\label{eq:FDE}
\end{equation}
to explain the current accelerated expansion of our Universe. Here $\rho$ can be composed of conventional matter, i.e, cold dark matter, baryons and radiation. For the origin of $\rho^n$ term, one can find several explanations \cite{ref:cardterm}. It can mimic the brane model which
include a power law term due to the embedding of our universe into a
five dimensional bulk. And, the late time accelerated expansion of our
Universe is because of the leakage of the gravity force at the large
scale in the brane world. The Cardassian model can reduce to $\Lambda$CDM
model when $n=0$ and to conventional CDM model with vanishing
cosmological constant when $A=0$. The new term $\rho^n$ will dominate the energy component at the late epoch to provide an accelerated expansion, then the values of $n$ should be $<2/3$.

This model has been confronted by cosmic observations extensively, when the energy density is composed of
cold dark matter and baryon, for the recent result please see \cite{ref:cardassiancon} and references therein, where $n=w+1=-0.039^{+0.135}_{-0.153}$ was obtained by using SN Ia, BAO, CMB shift parameters and observational Hubble data. In this case, the effective equation of state (EoS) of effective dark energy (the second term of Eq. (\ref{eq:FDE})) is $w^{eff}_{de}=n-1$. So it corresponds to quintessence dark energy model where the matter and dark energy are included only. However, it is not always true. When the radiation energy component is added in the Cardassian model, i.e. $\rho=\rho_m+\rho_r$, it is not equivalent to the quintessence model anymore due to the time variable effective EoS \cite{ref:Sen}
\begin{equation}
w^{eff}_{de}=(n-1)+\frac{n}{3}\frac{\rho_{r0}a^{-4}}{\rho_{m0}a^{-3}+\rho_{r0}a^{-4}}\label{eq:GEoS}
\end{equation}
where $\rho_0$ and $\rho_{m0}=\rho_{c0}+\rho_{b0}$ are the present energy density for radiation and matter and we have normalized the scale factor to $a_{0}=1$. At the early epoch after the recombination and before the last scattering surface, the radiation and the matter energy components are important, where the effective EoS $w^{eff}_{de}$ is not a constant. Therefore it is important to take the general form of EoS (\ref{eq:GEoS}) into account when one constrains Cardassian model from CMB observations. However, in the literature for example \cite{ref:cardassiancon}, this important thing was {\it not} considered at all, with exception of \cite{ref:Sen} (see also \cite{ref:spectrum}), where the locations of the peaks of the CMB anisotropic power spectrum were used as cosmic constraint. Though the time variable effective EoS was considered in \cite{ref:Sen}, only the positions of peaks of CMB power spectra were used as cosmic constraints. In that paper \cite{ref:Sen}, by fixing the values of $n_s$ and giving a number of values of $n$, the authors gave the ranges of $\Omega_{m0}$ and $h$. In fact, it is not enough to give a tight constraint to model parameter space. In Ref. \cite{ref:spectrum}, the authors investigated the CMB TT and matter power spectra. But, they did not give the model parameter space from the cosmic observations.

In the literature, the Cardassian model was constrained by the so-called CMB shift parameters, i.e. $R$, $l_a$ and $z_\ast$ which are obtained based on a $\Lambda$CDM model, from high redshifts. The problem that was almost overlooked is that the values of the CMB shift parameters depend on the cosmological models, here $\Lambda$CDM model. The potential logic is that since $\Lambda$CDM model is a concordance model then any  model which deviates slightly from $\Lambda$CDM model is a competitive model too. However it is non-proper when one uses cosmic observational data points to constrain any other cosmological models. Strictly speaking, it is just a test of the possible viability of a model \footnote{In the sense of small deviation from $\Lambda$CDM model.} not a constraint to model. This is because that the potential {\it circular problem} is committed and that is what we try to avoid in the cosmological constraint issue. In fact, the CMB shift parameters would be different for different cosmological models due to different physics process around the last scattering surface, for example early dark energy model \cite{ref:EDE} where the contribution from dark energy may not be neglected due to nontrivial equation of state (EoS) of dark energy. In this situation, it is dangerous to use data points derived in $\Lambda$CDM model due to much departure from $\Lambda$CDM model. One the other hand, the full CMB data points contain more information than the shift parameters. For instance, at late epoch when the dark energy is dominated, the gravitational potential becomes evolution. It affects to the anisotropy power spectra of CMB at large scale (low $l$ parts) due to the so-called integrated Sachs-Wolfe (ISW) effect which is sensitive to the properties of dark energy. Apparently, the CMB shift parameters do not include this information. So, one can expect a tight constraint to cosmological models when the full information for CMB is included. So, in this paper, we should take the observational data points from CMB directly not the derived model dependent CMB shift parameters.

To fill out the gap and to avoid the so-called circular problem, in this paper we shall use the type Ia supernova (SN Ia), baryon acoustic oscillations (BAO) and full WMAP-7yr data points to constrain the Cardassian model.

This paper is structured as follows. In section \ref{sec:method}, we give a brief review of Cardassian model and present the comic observational data sets and constraint methodalogy used in this paper. Section \ref{sec:conclusion} is the conclusion.

\section{Constraint Methodalogy and Results} \label{sec:method}

At first, we give a grief review of the Cardassian model. After the definition of an effective dark energy $\rho^{eff}_{de}$, the Friedmann equation (\ref{eq:FDE}) can be recast into
\begin{equation}
H^2=\frac{8\pi G}{3}\left(\rho+\rho^{eff}_{de}\right)
\end{equation}
where $\rho=\rho_{m0}a^{-3}+\rho_{r0}a^{-4}$ and $\rho^{eff}_{de}$ is given by
\begin{eqnarray}
\rho^{eff}_{de}&=&\frac{3A}{8\pi G}\rho^n\nonumber\\
&=&(\frac{8\pi G}{3H^2_0})^{1-n}\frac{(\Omega_{m0}+\Omega_{r0})^n}{(1-\Omega_{m0}-\Omega_{r0})}\rho^n
\end{eqnarray}
or equivalently
\begin{equation}
\rho^{eff}_{de}=\rho^{eff}_{de0}\left(\frac{\Omega_{m0}a^{-3}+\Omega_{r0}a^{-4}}{\Omega_{m0}+\Omega_{r0}}\right)^n
\end{equation}
where $\Omega_{m0}=\Omega_{c0}+\Omega_{b0}$ and $\Omega_{c0}=8\pi G\rho_{c0}/3H^2_0$
$\Omega_{b0}=8\pi G\rho_{b0}/3H^2_0$ and $\Omega_{r0}=8\pi G\rho_{r0}/3H^2_0$ are present dimensionless energy density for cold dark matter, baryon and radiation respectively. Here we have defined the dimensionless energy density for effective dark energy as $\Omega^{eff}_{de0}=8\pi G\rho^{eff}_{de0}/3H^2_0=1-\Omega_{m0}-\Omega_{r0}$ for spatially flat FRW Universe. One should notice that $n$ is the only model parameter.

We consider the perturbation equations for effective dark energy in a spatially flat FRW Universe. We treat the effective dark energy as a perfect fluid with EoS (\ref{eq:GEoS}). In the synchronous gauge, using the conservation of energy-momentum tensor $T^{\mu}_{\nu;\mu}=0$, one has the perturbation equations of density contrast and velocity divergence for effective dark energy
\begin{eqnarray}
\dot{\delta}_{de}&=&-(1+w_{de})(\theta_{de}+\frac{\dot{h}}{2})-3\mathcal{H}(c^{2}_{s}-w_{de})\delta_{de}\\
\dot{\theta}_{de}&=&-\mathcal{H}(1-3c^{2}_{s})\theta_{de}+\frac{c^{2}_{s}}{1+w}k^{2}\delta_{de}-k^{2}\sigma_{de}
\end{eqnarray}
following the notation of Ma and Bertschinger \cite{ref:MB}. For the perturbation theory in gauge ready formalism, please see \cite{ref:Hwang}. The shear perturbation $\sigma_{de}=0$ is assumed and the adiabatic initial conditions are adopted in our calculation.

In our analysis, we perform a global fitting to determine the
cosmological parameters using the Markov Chain Monte Carlo (MCMC)
method. We modified the publicly available {\bf cosmoMC} package \cite{ref:MCMC} to include effective dark energy in the CAMB \cite{ref:CAMB} code which is used to calculate the theoretical CMB power spectrum. The following $7$-dimensional parameter space  is adopted
\begin{equation}
P\equiv\{\omega_{b},\omega_c, \Theta_{S},\tau, n,n_{s},\log[10^{10}A_{s}]\}
\end{equation}
where $\omega_{b}=\Omega_{b}h^{2}$ and $\omega_{c}=\Omega_{c}h^{2}$ are the physical density of  baryon and cold dark matter respectively, $\Theta_{S}$ (multiplied by $100$) is the ration of the sound horizon and angular diameter distance, $\tau$ is the optical depth, $n$ is the newly added model parameter related to Cardassian model, $n_{s}$ is scalar spectral index, $A_{s}$ is the amplitude of the initial power spectrum. The pivot scale of the initial scalar power spectrum $k_{s0}=0.05\text{Mpc}^{-1}$ is used in this paper. The following flat priors to model parameters are adopted: $\omega_{b}\in[0.005,0.1]$, $\omega_{c}\in[0.01,0.99]$, $\Theta_{S}\in[0.5,10]$, $\tau\in[0.01,0.8]$, $n\in[-1,2/3]$, $n_{s}\in[0.5,1.5]$, $\log[10^{10}A_{s}]\in[2.7, 4]$. Furthermore, the hard coded prior on the comic age $10\text{Gyr}<t_{0}<\text{20Gyr}$ is also imposed. Also, the physical baryon density $\omega_{b}=0.022\pm0.002$ \cite{ref:bbn} from big bang nucleosynthesis and new Hubble constant $H_{0}=74.2\pm3.6\text{kms}^{-1}\text{Mpc}^{-1}$ \cite{ref:hubble} are adopted.

To get the distribution of parameters, we calculate the total likelihood $\mathcal{L} \propto e^{-\chi^{2}/2}$, where $\chi^{2}$ is given as
\begin{equation}
\chi^{2}=\chi^{2}_{CMB}+\chi^{2}_{BAO}+\chi^{2}_{SN}.
\end{equation}
The $557$ Union2 data \cite{ref:Union2} with systematic errors and BAO \cite{ref:BAO} are used to constrain the background evolution, for the detailed description please see Refs. \cite{ref:Xu}. SN Ia is used as standard candle. And BAO is used as standard ruler. To use the BAO information, we obtain the baryon drag epoch redshift $z_d$ numerically from the following integration \cite{ref:Hamann}
\begin{eqnarray}
\tau(\eta_d)&\equiv& \int_{\eta}^{\eta_0}d\eta'\dot{\tau}_d\nonumber\\
&=&\int_0^{z_d}dz\frac{d\eta}{da}\frac{x_e(z)\sigma_T}{R}=1
\end{eqnarray}
where $R=3\rho_{b}/4\rho_{\gamma}$, $\sigma_T$ is the Thomson cross-section and $x_e(z)$ is the fraction of free electrons. Then the sound horizon is
\begin{equation}
r_{s}(z_{d})=\int_{0}^{\eta(z_{d})}d\eta c_{s}(1+z).
\end{equation}
where $c_s=1/\sqrt{3(1+R)}$ is the sound speed. Also, to obtain unbiased parameter and error estimates, we use the substitution \cite{ref:Hamann}
\begin{equation}
d_z\rightarrow d_z\frac{\hat{r}_s(\tilde{z}_d)}{\hat{r}_s(z_d)}r_s(z_d),
\end{equation}
where $d_z=r_s(\tilde{z}_d)/D_V(z)$, $\hat{r}_s$ is evaluated for the fiducial cosmology of Ref. \cite{ref:BAO}, and $\tilde{z}_d$ is redshift of drag epoch obtained by using the fitting formula \cite{ref:EH} for the fiducial cosmology. Here $D_V(z)=[(1+z)^2D^2_Acz/H(z)]^{1/3}$ is the 'volume distance' with the angular diameter distance $D_A$. In this paper, for BAO information, the SDSS data points from \cite{ref:BAO} and the WiggleZ data points \cite{ref:wiggles} are used. For CMB data set, the temperature power spectrum from WMAP $7$-year data \cite{ref:wmap7} are employed to constrain the model parameters related to initial conditions.

\begin{center}
\begin{figure}[tbh]
\includegraphics[width=10cm]{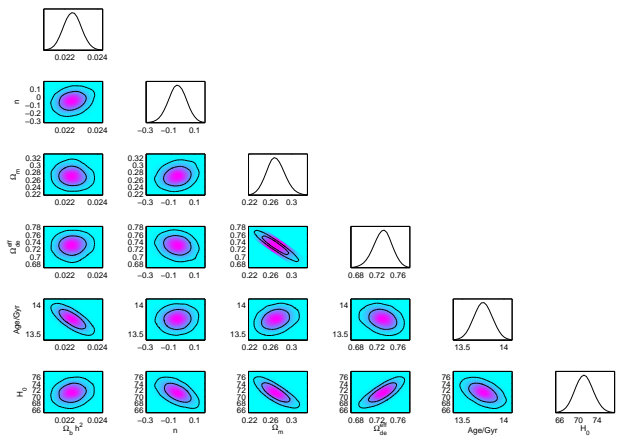}
\caption{The 2-D contours with $1\sigma$, $2\sigma$ regions and 1-D
marginalized distribution for Cardassian
model.}\label{fig:cont}
\end{figure}
\end{center}

The constrained results are shown in Tab. \ref{tab:results} and Fig.
\ref{fig:cont}. As a comparison to the result obtained in \cite{ref:cardassiancon} where $n=-0.039^{+0.135}_{-0.153}$ was obtained, a tight constraint was achieved in this paper apparently. To see the difference between the Cardassian and the quintessence model, we also redid the same process by using the same data sets combination, the obtained parameter space with $1,2\sigma$ regions is also listed in the last column of Tab. \ref{tab:results}.

\begingroup
\begin{table}[tbh]
\begin{center}
\begin{tabular}{ccc}
\hline\hline Prameters&Cardassian & Quintessence \\ \hline
$\Omega_b h^2$ & $    0.0225_{-    0.000525-    0.00102}^{+    0.000521+    0.00104}$  & $  0.0224_{-    0.000520-    0.00102}^{+    0.000517+    0.00103}$\\
$\Omega_{DM} h^2$ & $    0.114_{-    0.00453-    0.00872}^{+    0.00449+    0.00877}$ & $    0.114_{-    0.00441-    0.00859}^{+    0.00444+    0.00885}$ \\
$\theta$ & $    1.0391_{-    0.00256-    0.00507}^{+    0.00255+    0.00502}$ & $    1.0390_{-    0.00262-    0.00520}^{+    0.00260+    0.00518}$ \\
$\tau$ & $    0.0868_{-    0.00698-    0.0233}^{+    0.00636+    0.0242}$ & $    0.0868_{-    0.00719-    0.0230}^{+    0.00625+    0.0250}$ \\
$n(w)$ & $   -0.0479_{-    0.0732-    0.148}^{+    0.0730+    0.142}$ & $   -1.0492_{-    0.0751-    0.155}^{+    0.0745+    0.142}$\\
$n_s$ & $    0.966_{-    0.0128-    0.0252}^{+    0.0127+    0.0256}$ & $    0.966_{-    0.0126-    0.0248}^{+    0.0126+    0.0254}$ \\
$\log[10^{10} A_s]$ & $    3.0875_{-    0.0338-    0.0652}^{+    0.0338+    0.0681}$  & $    3.0877_{-    0.0337-    0.0657}^{+    0.0335+    0.0691}$ \\
$\Omega^{eff}_{de}$($\Omega_\Lambda$) & $    0.731_{-    0.0168-    0.0351}^{+    0.0169+    0.0320}$ & $    0.730_{-    0.0168-    0.0344}^{+    0.0167+    0.0311}$\\
$Age/Gyr$ & $   13.749_{-    0.109-    0.214}^{+    0.109+    0.211}$ & $   13.756_{-    0.109-    0.219}^{+    0.108+    0.219}$\\
$\Omega_m$ & $    0.269_{-    0.0169-    0.0320}^{+    0.0168+    0.0351}$ & $    0.270_{-    0.0167-    0.0311}^{+    0.0168+    0.0344}$\\
$z_{re}$ & $   10.524_{-    1.206-    2.400}^{+    1.204+    2.346}$ & $   10.538_{-    1.186-    2.390}^{+    1.1869+    2.447}$\\
$H_0$ & $   71.193_{-    1.864-    3.607}^{+    1.863+    3.727}$ & $   71.110_{-    1.877-    3.667}^{+    1.860+    3.745}$\\
\hline\hline
\end{tabular}
\caption{The mean values of model parameters for Cardassian and quintessence models with $1\sigma$ and $2\sigma$ errors, where WMAP $7$-year, SN Union2 and BAO data sets are used.}\label{tab:results}
\end{center}
\end{table}
\endgroup

\begin{center}
\begin{figure}[tbh]
\includegraphics[width=9cm]{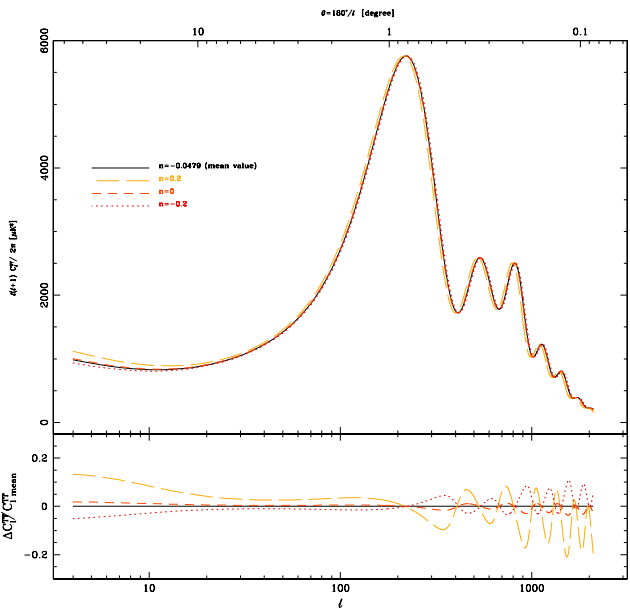}
\caption{The effect of model parameter $n$ to CMB power spectra, where the solid black lines for the mean values taken from Tab. \ref{tab:results}. The dashed and dotted lines are for the cases of different values of $n$ where the other relevant parameters are fixed to their mean values as listed in Tab. \ref{tab:results}. The bottom panel shows the corresponding ratios to the mean value case, i.e. $(C^{TT}_l-C^{TT}_{l \quad mean})/C^{TT}_{l \quad mean}$.}\label{fig:cls}
\end{figure}
\end{center}

To understand the effect of model parameter $n$ to CMB power spectra, we plot the Fig. \ref{fig:cls} by taking different values of $n$ where the other relevant parameter are fixed to their mean values as listed in the second column of Tab. \ref{tab:results}. Large values of model parameter $n$ increase and depress the CMB power spectra in the left and right sides of the first peak respectively. It is on the contrary for small values of $n$. For the fixed current ratio $\Omega^{eff}_{de0}/\Omega_{m0}$, a larger value of $n$ makes the effective dark energy dominate earlier, then the gravitational potential decays more, which enhance the lower $l$'s part of CMB power spectra due to Integrated Sachs-Wolfe (ISW) effect. One can also read off that large values of $n$ shift the position of peaks to the low $l$'s part slightly as shown in the bottom panel in Fig. \ref{fig:cls}, which comes from the change of the ratio of sound horizons at current and last scattering surface due to different values of EoS of effective dark energy.

\begin{center}
\begin{figure}[tbh]
\includegraphics[width=9cm]{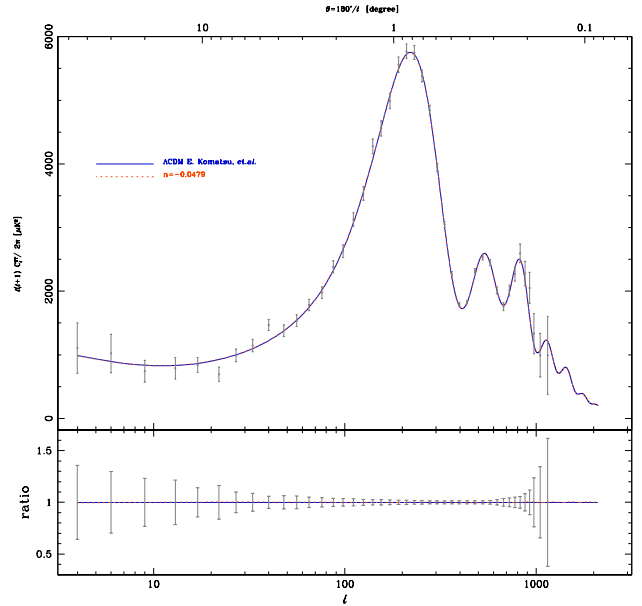}
\caption{The CMB $C^{TT}_l$ power spectrum v.s. multiple moment $l$, where the grey dots with error
bars denote the observed data with their corresponding uncertainties from WMAP $7$-year results, the red dashed lines are for the Cardassian model with mean values as shown in Table \ref{tab:results}, the blue solid lines are for $\Lambda$CDM model with mean values taken from \cite{ref:wmap7} with WMAP+BAO+$H_0$ constraint results. The bottom panel shows the ratios to $\Lambda$CDM model.}\label{fig:mean}
\end{figure}
\end{center}

Correspondingly, we show the comparison to $\Lambda$CDM model in Fig. \ref{fig:mean} where the mean values are taken from the second column of Tab. \ref{tab:results}. From Fig. \ref{fig:mean}, one can find the Cardassian model match $\Lambda$CDM model very well, and it almost in the center of error bars of WMAP-7yr data points. It means that current cosmic observational data points cannot discriminate Cardassian model from $\Lambda$CDM model.

\begin{center}
\begin{figure}[tbh]
\includegraphics[width=9cm]{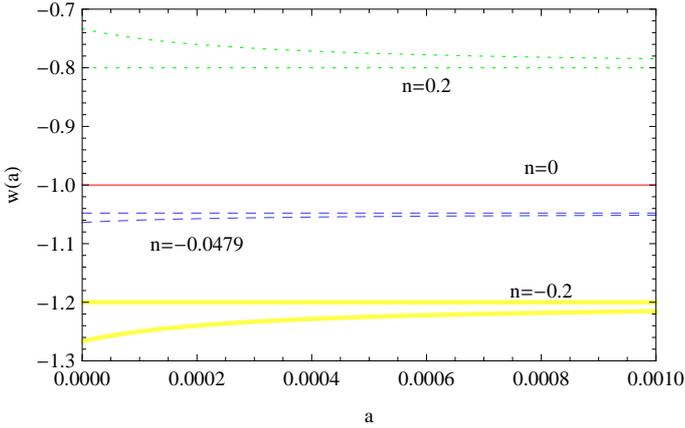}
\caption{The evolutions of EoS of effective dark energy, where different values of model parameter $n$ are adopted and the other relevant parameters are fixed to their mean values as listed in Tab. \ref{tab:results}.}\label{fig:eos}
\end{figure}
\end{center}

Now, we are in the position to see the effect to CMB power spectra due to the evolution of effective EoS at early epoch. At first, the evolutions of EoS of effective dark energy are shown in Fig. \ref{fig:eos} where different values of model parameter $n$ are adopted and the other relevant parameters are fixed to their mean values as listed in Tab. \ref{tab:results}. One can see the differences of the evolutions of EoS with respect to scale factor $a$ at early epoch apparently. And we expect to see the effects on CMB power spectra.

\begin{center}
\begin{figure}[tbh]
\includegraphics[width=9cm]{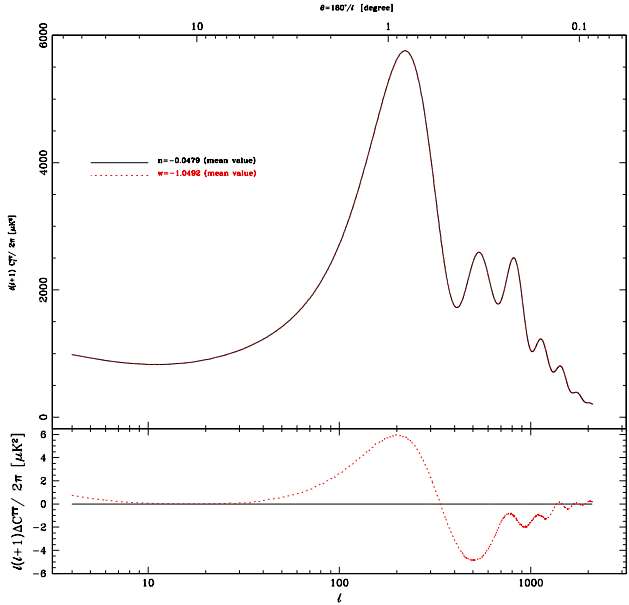}
\caption{The CMB power spectra for quintessence model and Cardassian model, where the values of model parameters are adopted as listed in Tab. \ref{tab:results}. The bottom panel shows the deviation from quintessence model.}\label{fig:compare}
\end{figure}
\end{center}

To do that we plotted the CMB power spectra for Cadassian and quintessence model in Fig. \ref{fig:compare} where the model parameters are fixed to their corresponding mean values as listed in Tab. \ref{tab:results}. From Fig. \ref{fig:compare}, one can easily see the differences between Carddssian and quintessence model that appear at high $l$'s part, i.e. the epoch around the last scattering surface as expected. However, the devotion is about the order $0.1/\%$ which almost cannot be discriminated by current data sets. But on theoretical level, they are really not equivalent.

\section{Conclusion}\label{sec:conclusion}

In summary, in this paper we have revisited the Cardassian model where the radiation energy component is included. In this situation, the EoS of effective dark energy for the Cardassian model is not a constant but time variable. Then the Cardassian model is not equivalent to quintessence model anymore. We performed a global fitting on the cosmological parameters in Cardassian model by using a completely consistent analysis where the full information from WMAP-7yr released data points were involved in a consistent way. We find out that the Cardassian model is consistent with current cosmic observational data sets. The constrained results are shown in Tab. \ref{tab:results}. The results show that the model parameter $n$ in $1,2\sigma$ regions are $n= -0.0479_{-    0.0732-    0.148}^{+    0.0730+    0.142}$ which are tighter than that, $n=-0.039^{+0.135}_{-0.153}$, obtained recently in \cite{ref:cardassiancon}. In Ref. \cite{ref:Sen}, the authors had used the positions of CMB peaks to constrain the Cardassian model. They found out that the values of $n$ depends on the values of $n_s$. They also pointed out that for $n_s=1$ the $n=0$ ($\Lambda$CDM model) is not included in the allowed region. In this paper, we have used the full information from WMAP-7yr results not the positions of peaks of CMB or the derived CMB shift parameters in $\Lambda$CDM model to constrain the model parameter space. The results show that $n=0$ ($\Lambda$CDM) case is included in the allowed region. It means that currently available data points from SN, BAO, HST and CMB cannot distinguish a Cardassian model from a $\Lambda$CDM model. We also show the effect of time variable EoS of effective dark energy to the CMB power spectra for the Cardassian model by comparison to that for the quintessence model. We find out the deviation is about the order of $\mathcal{O}(10^{-3})$ which cannot be discriminated by current CMB data sets.

\section{Acknowledgements} We thank an anonymous referee for helpful improvement of this paper. This work is supported by the Fundamental Research Funds for
the Central Universities (DUT10LK31) and (DUT11LK39) and in part by NSFC under Grants No. 11275035.

\end{document}